\documentclass[conference]{IEEEtran}
\ifCLASSINFOpdf
\else
\fi
\usepackage{cite}
\usepackage{booktabs}
\usepackage{hyperref}
\usepackage{graphicx}
\usepackage{xcolor}
\usepackage{xspace}
\usepackage{multirow}
\usepackage{rotating}
\usepackage{tcolorbox}

\begin{document}
%
\title{``These cameras are just like the Eye of Sauron'': A Sociotechnical Threat Model for AI-Driven Smart Home Devices as Perceived by UK-Based Domestic Workers}

\author{
Shijing He\IEEEauthorrefmark{1},
Yaxiong Lei\IEEEauthorrefmark{2},
Xiao Zhan\IEEEauthorrefmark{3},
Ruba Abu-Salma\IEEEauthorrefmark{1},
Jose Such\IEEEauthorrefmark{4}\\
\IEEEauthorrefmark{1}King's College London,
\IEEEauthorrefmark{2}University of St Andrews,
\IEEEauthorrefmark{3}Universitat Politècnica de València,
\IEEEauthorrefmark{4}INGENIO (CSIC-UPV)
}

%


\IEEEoverridecommandlockouts
\makeatletter\def\@IEEEpubidpullup{6.5\baselineskip}\makeatother
\IEEEpubid{\parbox{\columnwidth}{
		Symposium on Usable Security and Privacy (USEC) 2026 \\
		27 February 2026, San Diego, CA, USA \\
		ISBN 978-1-970672-07-7 \\
		https://dx.doi.org/10.14722/usec.2026.23030 \\
		www.ndss-symposium.org, https://www.usablesecurity.net/USEC/
}
\hspace{\columnsep}\makebox[\columnwidth]{}}

\maketitle

\begin{abstract}
The growing adoption of AI-driven smart home devices has introduced new privacy risks for domestic workers (DWs), who are frequently monitored in employers’ homes while also using smart devices in their own households. We conducted semi-structured interviews with 18 UK-based DWs and performed a human-centered threat modeling analysis of their experiences through the lens of Communication Privacy Management (CPM). Our findings extend existing threat models beyond abstract adversaries and single-household contexts by showing how AI analytics, residual data logs, and cross-household data flows shaped the privacy risks faced by participants. In employer-controlled homes, AI-enabled features and opaque, agency-mediated employment arrangements intensified surveillance and constrained participants’ ability to negotiate privacy boundaries. In their own homes, participants had greater control as device owners but still faced challenges, including gendered administrative roles, opaque AI functionalities, and uncertainty around data retention. We synthesize these insights into a sociotechnical threat model that identifies DW agencies as institutional adversaries and maps AI-driven privacy risks across interconnected households, and we outline social and practical implications for strengthening DW privacy and agency.
\end{abstract}

%

\section{Introduction}
The privacy implications of smart homes for bystanders (e.g., guests) are well documented (e.g.,~\cite{yao2019privacy,despres2024my,thakkar2022would,saqib2025bystander,abu-salma2025grand}). Prior research shows that power imbalances and misaligned preferences in multi-user smart homes can lead to privacy conflicts, excluding some household members from meaningful control over deployed devices~\cite{he2025exploring,bernd2022balancing,bernd2020bystanders,slupska2022they}. However, domestic workers (DWs) represent a distinct and more vulnerable class of bystanders. They are present in their employers’ smart homes not as guests but as employees, and their employers’ homes function as workplaces rather than spaces of leisure~\cite{bernd2020bystanders,albayaydh2022exploring,johnson2020beyond}. DWs include those who temporarily live in their employers’ homes to fulfil job responsibilities (e.g., au-pairs) and live-out DWs who spend limited time in employers’ homes~\cite{he2025exploring}. In both cases, DWs often lack bargaining power over how smart home devices are installed, configured, or monitored, even though these devices can capture intimate details of their daily routines.

Advanced AI capabilities are increasingly embedded in smart home devices. Systems with these capabilities offer personalized conversational experiences, persistent memory, and contextual data retention~\cite{daniel2025agent_ai}. These features can further shape DWs’ privacy perceptions and practices while working in employer-controlled smart homes~\cite{saqib2025bystander}. In this study, rather than covering the entire smart home ecosystem, we focus on AI-driven smart home devices and their inference capabilities enabled by machine-learning models that go beyond basic sensing or manual review (e.g., simply recording audio, or triggering notifications based on rule-based thresholds).

Specifically, we focus on two types of devices. First, AI-driven smart cameras are vision systems with built-in sensors that process visual information (e.g., detecting motion or recognizing a visitor’s face). These systems support automated detection and classification (e.g., person/pet/vehicle detection), event summarization and ``smart'' alerts, activity zones, and higher-level analytics such as crying detection or behavior-based inferences. Second, AI-driven smart speakers (also known as voice assistants) are controlled by spoken commands and use speech recognition and natural language processing to perform tasks such as information retrieval and device coordination. Advanced features include continuous or background voice processing, personalized responses, retention of interaction histories (e.g., voice logs), and, in newer systems, memory-like functions that enable persistent, context-aware personalization (e.g., Alexa+~\cite{alexa+}). These devices have been consistently identified as key sources of privacy risk~\cite{abu2023diverse,bernd2020bystanders,bernd2022balancing,slupska2022they,johnson2020beyond,albayaydh2022exploring,albayaydh2023examining,ju2023re,abu2024they,he2025exploring}. Accordingly, we focus on DWs’ interactions with these AI-driven devices, as their capabilities most directly shape perceptions of surveillance, inference, and control.

Prior studies have examined device-specific perceptions and perceived privacy threats in workplace contexts (see \S\ref{related_dw_privacy}). He et al.~\cite{he2025exploring} also investigated the privacy perceptions and practices of Chinese migrant DWs in their own homes, where they acted as device owners and administrators. In this context, DWs’ ``own homes'' referred to spaces where they had primary control over configuring and using smart devices, such as their main residence or a separate household they maintained (e.g., a family home they regularly visited or supported). However, existing literature overlooks how being surveilled by employer-owned, AI-driven smart home devices at work, while simultaneously acting as owners and administrators of personal devices at home, shapes DWs’ evolving privacy boundaries and threat perceptions. To address this gap, we ask:

\textit{RQ: How do DWs based in the UK experience and interpret AI-driven smart home devices in both employer-controlled and personal home contexts, and what socio-technical threats do they identify across these two contexts?}

We conducted 18 semi-structured interviews with DW participants residing in the UK to gain an in-depth understanding of how they navigate privacy in both employer-owned and personal smart homes. Drawing on recent calls to formalize human-centered threat modeling for at-risk users (see \S\ref{related_threat_model}) and employing the theoretical framework of Communication Privacy Management (CPM)~\cite{petronio1991communication} (see \S\ref{cpm}), we develop a threat model that considers context, threats, protective strategies, and reflections grounded in the lived experiences of (migrant) DWs in AI-driven smart homes—dimensions that prior work has not fully captured (e.g.,~\cite{slupska2022they,abu2024they}). We contribute a dedicated human-centered threat model grounded in DWs’ experiences of navigating two interconnected socio-technical contexts: their employers’ homes and their own. This model identifies (i) AI analytics and cross-household data traces as concrete threats and (ii) DW agencies and institutions as key adversaries shaping surveillance defaults.

Our findings demonstrate how AI-driven smart home devices create technical threat pathways (e.g., behavioral analytics) and ``orphaned'' DW profiles that may persist beyond employment. We show how surveillance at work and privacy management at home mutually reinforce one another, as DWs transfer skills, norms, and expectations across personal and employer-controlled smart home environments. Importantly, threat models cannot be confined to a single household; DWs’ threat perceptions and defenses are co-constructed across interconnected homes. We also identify an additional socio-technical privacy threat: agency-mediated practices, such as vague employment contracts, limited training, and employer-sided dispute handling, which further constrain DWs’ ability to contest surveillance or renegotiate privacy rules. Finally, we translate these insights into socio-technical recommendations for device vendors, platform providers, DW agencies, and policymakers aimed at reducing surveillance harms from AI-driven smart home devices and strengthening DWs’ capacity to manage privacy across interconnected households and contexts.

\section{Related Work}
\label{related}
\subsection{Privacy Concerns and Power Dynamics in Smart Homes}
\label{related_smart_home_privacy}

Prior work has broadly examined how information asymmetries and managerial control are reinforced through data-driven monitoring and platform governance, shaping what DWs can contest and what forms of resistance are feasible (e.g.,~\cite{rosenblat2016algorithmic,sum2025future,awumey2024systematic}). In particular, studies have shown that privacy in multi-user smart homes is tightly intertwined with relationship dynamics and unequal control over technologies. Smart home devices can support coordination and independence but can also enable surveillance and conflicts over use and data access~\cite{apthorpe2022you,he2018rethinking}. Residents’ coping strategies (e.g., avoidance versus resignation) differ depending on whether they are dealing with co-residents or external service providers~\cite{huang2020amazon}, and tensions often arise between primary users—who typically own, deploy, and configure devices—and other household members with limited input into how settings are configured~\cite{geeng2019s}. The person who installs a device generally sets the terms of its use, and perceptions of who should control technology are deeply entangled with power dynamics~\cite{ehrenberg2021technology}. Expectations about bystanders’ control also vary by device location, the device’s effect, and the relationship between primary users and bystanders~\cite{kraemer2020further}.

Research on smart home bystanders, such as visitors and guests, has further examined how privacy concerns are shaped by trust, consent, and perceived benefits. Bystanders’ data-sharing preferences depend in part on how much they trust the device owner as a data recipient~\cite{marky2020you}. For example, Stephenson et al.~\cite{stephenson2023s} identified stages and barriers that domestic violence victims and survivors face in smart homes when attempting to detect and mitigate abuse. These challenges often stem from device opacity, shared ownership, and weak legal and support infrastructures, demonstrating how smart home devices can entrench rather than alleviate unequal power dynamics in domestic spaces. Specifically, in the context of smart cameras, bystanders are often particularly concerned about video collection in others’ homes relative to their own, reflecting the asymmetry between those who benefit from the device and those who are merely exposed to it~\cite{despres2024my}.

\subsection{Privacy Concerns of (Migrant) DWs}
\label{related_dw_privacy}

DWs often face privacy challenges stemming from power imbalances and surveillance practices in employer-controlled smart home environments~\cite{bernd2020bystanders,bernd2022balancing,slupska2022they,albayaydh2022exploring,albayaydh2023examining,abu2024they,he2025exploring}. They are frequently present in the smart home and may occasionally co-use devices, yet typically lack ownership or configuration rights and are often treated as having little stake in the system~\cite{abu2024they}. Choe et al. found that primary users are particularly reluctant to disclose smart devices to DWs compared to family or friends~\cite{choe2012investigating}, highlighting sharper power imbalances in employer–employee relationships than in more egalitarian visitor–host relationships.

Smart home devices can significantly impact DWs, who often have limited control over privacy settings at work. For example, over 44\% of DWs in the UK report experiencing privacy issues at work~\cite{yilmaz2023exploring}, and some may even be viewed as ``security threats''~\cite{slupska2022they}. Bernd et al.~\cite{bernd2020bystanders,bernd2022balancing} highlighted power imbalances in which nannies often had little control over surveillance devices, emphasizing the importance of transparent communication and informed consent to build trust and improve working conditions. Abu-Salma et al. reported that DWs were generally comfortable with devices such as smart speakers and TVs but felt uneasy about indoor cameras, while employers prioritized household safety over workers’ privacy~\cite{abu2024they}.

Several studies have examined DWs’ privacy concerns in non-Western contexts. For example, Albayaydh et al. highlighted how smart home benefits can be undermined by power imbalances, calling for legal protections for migrant DWs in Jordan~\cite{albayaydh2022exploring,albayaydh2023examining,albayaydh2024innovative,albayaydh2024co}. In Hong Kong, Johnson et al. found that surveillance intended to ensure safety can erode DWs’ trust and affect care quality~\cite{johnson2020beyond}. Ju et al. reported that DWs were more accepting of nanny cams during the pandemic, particularly when they had positive relationships with employers~\cite{ju2023re}. He et al. showed that Confucian values, legal gaps, and power imbalances shape privacy risks in Chinese smart homes~\cite{he2025exploring}.

Furthermore, the immigration status of DWs can heighten privacy and security risks because workers may face greater dependency on employers (and sometimes DW agencies) for income, accommodation, and continued employment, alongside language barriers, limited access to rights information, and fear of retaliation~\cite{bernd2020bystanders,johnson2020beyond}. Slupska et al. found that UK migrant DWs prioritized protecting against threats such as government surveillance and online harassment over monitoring by employers, calling for tailored privacy protections~\cite{slupska2022they}.

However, most existing work has either treated DWs solely as bystanders in employer-controlled homes or mapped threats at a macro level without examining device ecologies or tracing DWs’ experiences across contexts. While He et al. explored DWs’ privacy in their own homes and noted agency-related issues~\cite{he2025exploring}, they did not examine how roles and expectations compare and contrast between home and work, distinguish AI-driven devices from traditional ones, or view agencies as active threat actors. Our study addresses these gaps by showing how AI capabilities of smart devices reshape DWs’ understanding of what data is collected, who can access it, and when it is deleted or leaves the smart home ecosystem. In employer homes, devices are configured by others in employer homes, whereas in their own homes, DWs decide whether to purchase, deploy, disable, or avoid them.

\subsection{Threat Modeling for (Migrant) DWs}
\label{related_threat_model}

Mainstream security and privacy threat models are typically designed around an average user, overlooking at-risk populations such as DWs, whose digital safety risks are shaped more by structural conditions than individual choices. Warford et al. proposed ten contextual risk factors, including legal–political context and marginalization, to characterize at-risk users’ heightened exposure to harm~\cite{warford2022sok}. Bellini et al. emphasized that at-risk users face a higher likelihood of being targeted and offered guidance for conducting safer research with these populations~\cite{bellini2024sok}. Usman and Zappala defined human-centered threat modeling as understanding how people are vulnerable to harm and what strategies they use to stay safe, proposing a process-oriented framework for eliciting mental models, harms, and coping strategies~\cite{usman2025sok}. These works highlight structural risk and the importance of centering lived practices but remain largely population-agnostic, providing methodological guidance rather than domain-specific threat models for labor-mediated smart home surveillance or cross-household dynamics.

Specifically, Slupska et al. identified government surveillance, online harassment, and employer monitoring as key threats, describing harms that span both online and offline domains~\cite{slupska2022they}. Abu-Salma et al. developed device-specific, multi-party threat models, showing that DWs view employers as primary threats with regard to in-home cameras and have limited control over devices and data flows~\cite{abu2024they}. However, both studies focused on employer-controlled smart homes, DWs were largely treated as bystanders, and emerging AI capabilities, such as behavioral analytics and long-term memory, were not central to the threat landscape.

While these studies demonstrate that DWs face structurally-shaped digital safety risks and that human-centered threat modeling must foreground practices, coping strategies, and power relations, existing models do not account for DWs’ dual positionality: (i) as bystanders in employer-controlled, AI-driven smart home environments, and (ii) as owners or administrators of devices in their own homes. Further, these models do not explain how AI-driven devices mediate harms across interconnected households, or how DW agencies contribute to or amplify these risks. In contrast, our study examines DWs’ experiences in both employer and personal smart home environments, investigating how their dual roles shape privacy boundary negotiations and how threat perceptions and mitigation strategies travel between contexts.

\section{Theoretical Framework}
\label{cpm}

We grounded our analysis in Communication Privacy Management (CPM) theory~\cite{petronio1991communication,petronio2017communication,petronio2020conceptualization}, using it both to interpret DWs’ privacy boundaries and to structure our socio-technical threat model. CPM explains how individuals manage private information in interpersonal settings along three dimensions: (1) \emph{privacy ownership}, how individuals see themselves (or others) as legitimate owners and controllers of personal information; (2) \emph{privacy control}, the rules they use to regulate information flows based on norms, preferences, and context; and (3) \emph{privacy turbulence}, what occurs when those rules are violated and boundaries must be renegotiated.

We used these components when developing our threat model for AI-driven smart home devices as experienced by DWs. In this model, DWs’ private information and the privacy rules they attempt to enforce constitute core \emph{assets}, boundary violations and unauthorized disclosures represent primary \emph{harms}, structural power imbalances with employers, agencies, and technically privileged family members form central \emph{vulnerabilities}, and episodes of privacy turbulence capture the lived \emph{impact} of these dynamics. This CPM-grounded perspective expands prior DW threat models by making boundary work itself a first-class consideration: it reveals who is treated as a legitimate ``owner'' of data in employer-controlled versus own-home ecologies, how co-ownership, e.g., agency contracts, family-administered devices, shifts control away from DWs, and how AI capabilities, e.g., crying detection, behavior analytics, persistent voice logs, create new pathways for turbulence. It also foregrounds DW agencies as institutional intermediaries whose contractual, informational, and training practices can normalize or amplify surveillance from AI-driven smart home devices, while tracing how residual risks and harms propagate between workplaces and personal homes.
\section{Methods}
\subsection{Participant Recruitment}
\label{participant_recruitment}

We recruited DW participants through Facebook groups and our university mailing lists. Participants were recruited based on a clearly defined inclusion criterion: regular interaction with at least one AI-driven smart home device in both contexts—their employers’ smart homes and their own homes. Although some live-in DW participants resided in their employers’ homes, these spaces also functioned as workplaces in which employers controlled devices and defined surveillance norms~\cite{bernd2020bystanders}. In contrast, DWs’ ``own-home'' contexts refer to environments where they or their families primarily managed devices and privacy rules. Live-in DW participants described navigating different boundaries around communication, rest, and daily routines in employer-controlled spaces compared to the autonomy they exercised over devices in their own homes.

To verify eligibility, all interested individuals first completed an online screening survey that asked about their regular use of devices such as surveillance cameras, smart doorbells, and smart speakers, particularly those with AI functions, in both employer-controlled and personal contexts. While this screening confirmed that all participants met the inclusion criterion, it also revealed a common asymmetry: participants frequently encountered more advanced AI capabilities (especially in surveillance cameras) in their employers’ homes, whereas their own smart devices often featured fewer or less complex AI functions. Our goal was to recruit a diverse sample of DW participants in the UK, including individuals with varied educational backgrounds, ethnicities, ages, and genders. Eligible roles included, but were not limited to, nannies, babysitters, caregivers, and house cleaners, aged 18 or older, who had experience working in an employer’s smart home and who also used smart home devices in their own homes (see participant demographics in Table~\ref{tab:demographic}). In line with UK compensation norms~\cite{uksalary}, each participant received a £20 gift card as a token of appreciation for their time and input.\footnote{The screening survey, interview guide, exit survey, and codebook are available in \href{https://osf.io/txh7c/overview}{\textbf{this OSF repository}}.
}

\subsection{Interview Procedure}
\label{interview_procedure}

We conducted online interviews with participants via Microsoft Teams, which also recorded and transcribed each session. Interviews lasted an average of 74.1 minutes. Three authors iteratively analyzed the data after each interview and observed that insights began to repeat, with no new insights emerging~\cite{guest2006many}. Data saturation was reached after the 16th interview; however, we conducted two additional interviews to confirm this assessment. While the concept of data saturation in qualitative research is debated~\cite{braun2006using} and subjective~\cite{braun2021saturate}, we adopted a reflexive approach by continuously reviewing our notes throughout data collection. Our final sample size aligns with qualitative research principles that emphasize depth and richness of insights over statistical generalizability~\cite{vasileiou2018characterising}.

We began by asking participants about their work experience in the UK, followed by their definitions of privacy. Participants then described their experiences with smart home devices in both employer-controlled and personal contexts. For each device, we asked about its location and data practices (e.g., types of data collected, access permissions, data storage), and probed into perceived differences in usage and placement across the two settings. We then explored experiences with smart home devices, particularly cameras and speakers featuring advanced capabilities, including how participants learned about device functionalities (e.g., behavioral analysis) and negotiated camera use. We also discussed participants’ concerns about inappropriate data disclosure and retention, as well as their emotional responses.

Because participants’ understanding of ``AI features'' varied, we used capability-focused probes rather than relying solely on the term ``AI'' (e.g., asking whether a device classified people or activities, or inferred events such as crying). We next examined the strategies participants used to manage privacy across contexts, including conflicts with DW agencies over employment contracts, training, and device settings. Participants reflected on household surveillance, autonomy in setting privacy rules, and differences in privacy control at work versus at home. We also explored how participants responded to privacy breaches (e.g., unauthorized access) and their perceived ability to act. Finally, participants shared advice for peers, suggested design ideas for shared device control, and reflected on differences related to laws and regulations between work and home settings.

\textbf{Pilot Interviews.}
We conducted two pilot interviews—one with a colleague and another with a DW (a UK citizen) who had previously worked as a live-in nanny—to refine the interview guide. Although these interviews were not included in the analysis, they helped strengthen sections on participants’ legal and regulatory knowledge and its influence on privacy perceptions, their expectations of DW agencies, and their experiences with AI-driven smart cameras and smart speakers.
\subsection{Data Analysis}
\label{method_data_analysis}

We first coded the interview transcripts inductively. Three authors independently coded the same four randomly selected transcripts and developed individual codebooks, yielding 117 initial codes per author. The authors then met to discuss discrepancies and consolidate these into a unified codebook comprising 84 codes. Using this merged codebook, they independently coded an additional transcript (distinct from the initial four) to test its consistency. Following further discussion and refinement, the revised codebook was reduced to 76 codes. This process was repeated twice more, and the codebook stabilized at 66 codes when code saturation was reached. Using the finalized codebook, the first author coded all 18 transcripts, while the other two authors each independently coded nine transcripts (Authors 1 and 2 coded P1–P9; Authors 1 and 3 coded P10–P18). Inter-rater reliability (IRR) was then measured, yielding an average $\kappa = 0.794$ for Authors 1 and 2 and an average $\kappa = 0.813$ for Authors 1 and 3, indicating excellent agreement~\cite{fleiss2013statistical}. The authors then organized the codes into higher-level themes and mapped them onto the CPM framework to contextualize how participants navigated privacy boundaries in both employer-controlled and personal contexts.

\subsection{Ethical Considerations}
\label{method_ethics}

In line with Bellini et al.’s recommendations~\cite{bellini2024sok}, our recruitment and interview procedures were designed to minimize risk, avoid dependency on employers or DW agencies, and provide participants with clear consent and exit options. We implemented several data protection safeguards during analysis. First, we removed all identifying information about participants, including real names, residential addresses, employers’ names, and other identifying characteristics. Second, we used pseudonyms for all participants and included only anonymized quotes in the paper. Third, access to the data, including interview transcripts and supporting documents such as signed consent forms and information sheets, was restricted to the authors. This study was reviewed and approved by the Research Ethics Committee at King’s College London.

\subsection{Limitations}
\label{method_limitation}
As with all qualitative research, our findings rely on self-reported data, which might be influenced by social desirability bias~\cite{nederhof1985methods}. To mitigate this, we used open-ended, neutrally phrased questions, built rapport through repeated assurances of confidentiality, and encouraged concrete examples. Our findings reflect experienced privacy boundaries and disruptions rather than direct measurements of technical setups or employer practices. Participants’ dual roles might have shaped their self-presentation: as vulnerable in employer-controlled homes, yet responsible and privacy-aware in their own homes, which could reinforce self-perceptions rather than fully reflect actual practices~\cite{bossauer2020trust}. We interpreted these accounts cautiously, noting that perceived control at home might not always align with privacy-preserving configurations. Participants typically encountered more advanced AI features in employer-controlled homes, which may explain differences in perceived risk. Our analysis focused on AI-enabled smart cameras and speakers, identified as the most intrusive devices, and we did not claim generalizability to all smart home technologies. We also did not interview employers or DW agencies, nor inspect devices directly; the threat model is based solely on participants’ reports. Finally, all materials were in English, which may have constrained some migrant participants’ articulation of technical or legal concepts.

\section{Results} \label{results} We present the key threats participants faced in their employers' smart homes, including pervasive camera surveillance (\S\ref{pervasive_surveillance}), risks posed by AI-driven smart home devices (\S\ref{AI_threat}), and threats originating from DW agencies (\S\ref{agency-threat}). We also examine how participants exercised privacy ownership/control in their own homes (\S\ref{own_home}) and how experiences in one context (employers’ homes) informed and reshaped their privacy expectations and practices in another (own homes) (\S\ref{work_home_reshape}).

\subsection{Pervasive Camera Surveillance} \label{pervasive_surveillance}
In employers’ smart homes, participants primarily described themselves as bystanders with little to no control over their privacy, frequently subjected to unauthorized surveillance. All participants expressed concern about smart cameras collecting their personally identifiable information (PII), monitoring their work performance, and intruding into their everyday lives. For example, P1, a Filipino live-in nanny, described her previous working conditions as a migrant DW, highlighting pervasive surveillance and a lack of privacy, including cameras installed in her bedroom. Several participants likened the experience to being watched from every angle, noting that the sheer number and placement of cameras left them with no space to simply be themselves. As P4 described: \textit{“These cameras are just like the Eye of Sauron.”} P1 further reflected on the mental health impact of camera monitoring combined with physical mistreatment: \textit{“I don't have a choice. They treated me like an animal [...] I want to know why they need cameras everywhere? If they can't expose these, I have PTSD from it, and I won't work in a house with cameras.”}

\textbf{Keeping Clear Privacy Boundaries was Key.} Many participants emphasized the broader concept of private boundaries, highlighting that safeguarding privacy was essential and that unauthorized surveillance constituted a boundary violation and breach of trust. P4 stressed the importance of negotiating the presence and use of monitoring devices: \textit{“We need to have a certain level of privacy, like a privacy zone or break area, that their monitoring devices cannot infringe upon without our consent.”} Several participants also noted that disclosing the presence and use of smart home cameras was crucial to prevent \textit{“spylike monitoring.”} For instance, P5 expressed the need for transparency in surveillance practices to clarify privacy boundaries: \textit{“Sharing footage with others goes way beyond what I'm comfortable with privacy-wise. It would be great if they could explain what they're monitoring and why. Knowing this would help me feel respected.”}

Yet, some participants noted that they were willing to accept limited sharing of certain personal attributes (e.g., faces) with their employers’ family members for benign purposes, such as entertainment or demonstrating work performance. However, many viewed sharing surveillance footage on social media (e.g., TikTok) without consent as a clear violation of established norms around privacy and trust. As P5 stated, \textit{“I will say no. That would definitely cross a line for me. It's one thing to monitor for security purposes, but broadcasting my actions to the world without my consent feels like a violation of privacy.”} Others described a constant sense of self-consciousness under pervasive monitoring. P15 noted, \textit{“You start to act for the camera, not for yourself,”} capturing a panoptic effect in which the possibility of being watched shaped participants’ behavior even when no one was visibly present.

\textbf{Surveillance Offered Benefits.} While participants were concerned about the lack of privacy and constant monitoring, many also acknowledged the benefits of smart cameras. These included demonstrating their work performance, enhancing care quality for children and the elderly, and providing evidence in case of conflicts or disputes. For example, P7 highlighted the value of cameras in clarifying disagreements: \textit{“I believe there could potentially be benefits. If I recall a situation one way and the child remembers it differently, the presence of the cameras could help clarify what actually occurred, providing a form of proof for both parties involved.”} A few participants also reported feeling relatively secure and safe in their employers’ homes, particularly when monitoring was governed by explicit agreements and clear communication. They recognized that their employers—and in some cases security staff—were actors in the system and involved in data flows, and they accepted certain forms of surveillance (e.g., hallway footage) as necessary for safety. For instance, P2, who worked for a government official with bodyguards and over 20 smart cameras, described the extensive camera deployment as providing a sense of being \textit{“100\% secure.”}

\begin{tcolorbox}[colback=green!10, colframe=black!20, boxrule=0pt, sharp corners=all]
\textbf{Key Takeaways:} In employer-controlled homes, participants viewed surveillance as pervasive and intrusive—cameras were often used without meaningful consent—creating a panoptic effect. Yet, participants acknowledged that cameras could enhance safety if used transparently and within defined boundaries.\end{tcolorbox}

\subsection{Risks Posed by AI-Driven Smart Home Devices} \label{AI_threat}
In their employers’ homes, participants perceived AI-driven smart home devices not merely as passive recorders but as systems capable of interpreting, storing, and later resurfacing their behavior, shaping how they were evaluated as DWs. They distinguished these threats from ordinary monitoring, emphasizing the role of automated judgment and the enduring nature of data traces left by AI-driven smart cameras and smart speakers. Although participants acknowledged the security and safety motivations behind deploying AI-driven smart cameras, some live-in participants specifically noted that, compared to standard cameras, those with features such as crying detection increased their privacy concerns and workplace discomfort. They were concerned that these interpretations could be used as de facto performance metrics—for example, how quickly they responded when the system flagged motion or crying—and that such logs might be consulted in disputes about care quality or diligence, even when participants had no control over how alerts were configured. For example, P15 explained how AI-driven smart cameras affected her emotional well-being, stating: \textit{“It felt kind of intrusive to be under constant surveillance. The camera can detect motions and even baby crying. Once, the camera just rotated directly toward me when I was holding the crying baby [...] such monitoring made me feel like I had almost no personal space.”}

Some participants emphasized that AI-driven smart cameras made privacy boundaries more opaque, as these devices could not only monitor visible actions but also interpret behaviors, emotions, and routines, often without participants’ awareness or consent. They questioned the validity of such algorithmic interpretations and whether privacy could meaningfully exist in environments shaped by intelligent surveillance technologies. P15 noted: \textit{“It's not just a camera anymore. It feels like it's watching what kind of person I am, not just what I do.”}

\textbf{Residual Voice and Data Traces in AI-Driven Smart Speakers.} Interestingly, we found that participants’ encounters with traces of previous workers’ data collected and stored by their employers’ smart speakers (e.g., stored voice histories or records) heightened their awareness of limited privacy control. This data persisted beyond the end of employment and remained accessible to employers and platforms, raising fears that participants’ own interactions could similarly be replayed, scrutinized, or repurposed long after they had left the household. Participants expressed particular concern about AI-driven smart speakers with memory functions in employer-controlled homes, where surveillance and data collection were perceived as extending beyond job-related oversight into deeply personal territory. A few participants emphasized that retaining orphaned data—information left behind by previous DWs—clearly crossed privacy boundaries, representing a loss of control for former workers and raising concern that the opaque nature of these systems could similarly undermine their own privacy. P15 shared a troubling experience involving leftover data from a previous DW: \textit{“It made me uneasy; if her name is still stored, what about mine? I felt exposed and worried the system might also be saving my habits and routines, leaving my data behind after I leave.”}

Some participants (e.g., P16, P18) also raised concerns about consent and transparency regarding data transmission. They also underscored the ethical importance of managing data post-employment. P18 noted: \textit{“It's unfair. If the relationship ends, all personal behavioral records should be automatically removed. Employers should adopt clear and transparent policies. We should know exactly what data is retained and have the right to consent or refuse.”}

\textbf{Blurred Professional Boundaries.} A few participants noted that some employers asked their DWs to monitor their children’s (mostly aged 6–12) interactions with AI-driven smart speakers (e.g., storing children’s prompts). This raised concerns about caregiving responsibilities being expanded to include surveillance tasks. Participants thus found themselves in a double bind: expected to act both as caregivers and as informal content moderators or data stewards. While reviewing children’s AI interactions, they often felt that their monitoring responsibilities extended to them, implicating their own privacy, exposing them to potential blame if something went wrong, and increasing pressure to always behave correctly in front of employers and devices. P15 described the complexity of managing boundaries, recounting a situation in which she reviewed a child’s AI assistant queries: \textit{“It felt like a privacy intrusion for both me and the child.”} P7 argued that assigning moderation tasks without clear guidelines blurred professional boundaries, infringing upon DWs’ privacy expectations and undermining clarity about who controlled such information.

\textbf{Threats from AI Functions and Intensified Privacy Turbulence.} We found that AI functions (e.g., (mis)interpretations, data retention) not only extended existing forms of surveillance but also deepened participants’ sense of being constantly judged, recorded, and potentially punished on the basis of digital traces they neither controlled nor fully understood. These AI functions intensified pre-existing power imbalances between employers and employees (DWs) and heightened privacy turbulence. Many participants described how surveillance in general, and AI-enabled functions in particular, restricted their autonomy and decision-making at work. Knowing that cameras and logs could be reviewed at any time contributed to a perception that any misstep—such as missing an alert, taking a break, or moving out of frame—might be interpreted as laziness or negligence, even when participants were already overburdened. For example, P10 described losing autonomy early in her caregiving career due to the combined pressure of heavy workloads and constant monitoring by AI-driven smart cameras with motion-detection features.

Some participants highlighted how constant monitoring reshaped their identity and sense of self in the workplace. P4 explained how being under surveillance shaped their actions and expectations, leading them to perform in ways that conformed to perceived norms rather than behaving naturally: \textit{“It's like always having a social spotlight on you—you’re not entirely yourself [...] there’s a sense of self-consciousness that comes with being monitored.”} In more severe cases, these technologies operated within already coercive environments, amplifying participants’ exposure to privacy and safety risks. P1 recounted extreme power abuses and described how fear overrode her privacy boundaries: \textit{“I couldn't refuse demands because I saw a colleague get slapped after our employer saw her using her phone through the camera and then demanded to check it. When it happened to me, I gave mine over right away. Fear of physical harm made me comply [...] Now AI is watching everything.”}

\begin{tcolorbox}[colback=green!10, colframe=black!20, boxrule=0pt, sharp corners=all]
\textbf{Key Takeaways:} AI-driven smart home devices introduced new risks in participants’ workplaces. Automated analytics and persistent logs made them feel constantly monitored, while tasks like overseeing children’s use of smart devices blurred caregiving and surveillance roles. These dynamics amplified power imbalances, turning AI devices into tools for documentation, justification, and control.
\end{tcolorbox}

\subsection{Threats Posed by DW Agencies} \label{agency-threat}

\textbf{Limited Assistance for Migrant DWs.} Some participants described how the dynamics of employer–agency relationships created an additional layer of surveillance and control, intensifying privacy turbulence. For instance, P8 noted: \textit{``Many agencies not only charge high agent fees, like 20\% to 30\% of our first-month salary, they also focus on how to maximize their profit from us and employers. They mostly prioritize their clients' interests, always listen to clients' complaints, but rarely consider ours.''} Some participants further highlighted that their DW agencies provided no training or guidance on managing privacy, underscoring a lack of transparent policies governing data flows and protection. As AI-driven smart home devices became increasingly embedded in these households, this silence and one-sided alignment effectively normalized sophisticated monitoring as a taken-for-granted part of the job—without equipping DWs with the knowledge or training necessary to understand or contest the specific risks of AI functions. P3 expressed frustration with this lack of structural support: \textit{``They said such training [on privacy and protecting against surveillance] was not part of their service. They didn't provide any support or training on how to handle privacy issues.''}

These dynamics were particularly acute for migrant DWs recruited from abroad~\cite{roberts2020rights}, for whom the agencies functioned not only as information gatekeepers but also as gatekeepers of residence and income. Several participants, including P1 and P15, described feeling unable to refuse placements with pervasive or intrusive surveillance—or to challenge employers' use of smart cameras and speakers—because they feared losing their job, jeopardizing their visa, or being ``blacklisted'' for future placements. In such cases, the agency's alignment with employers' interests, combined with the lack of privacy support, amplified the coercive potential of AI-enabled monitoring, rendering DWs' consent to surveillance largely nominal.

\textbf{Unclear Employment Contracts Amplified AI-Driven Risks.} Many participants\footnote{In the UK, an employer is required to provide a worker with a written statement of the main terms of their employment contract~\cite{ahlberg2022the}.} raised serious concerns about the transparency and clarity of employment contracts, particularly regarding privacy and monitoring practices. They emphasized the need for explicit privacy clauses specifying what information (e.g., video footage or behavioral logs) could be collected, how such data would be used, and under what conditions. In the absence of these details, participants often entered workplaces without a clear understanding of surveillance boundaries. For example, P1 described discrepancies between contractual promises and actual working conditions, illustrating how expectations of transparency and informed consent were not upheld: \textit{``The initial contract said I have a day off once a month and would receive £309\footnote{Due to ethical considerations, we converted the original currency to GBP based on the exchange rate at the time.} per month. It also stated the employer would provide shampoo and toothpaste. But this never happened, and the contract didn't state any cameras in the house.''}

DW agencies did not merely coexist with surveillance or offer contractual guidance; they actively shaped how AI-enabled monitoring became normalized and difficult to contest. Several participants raised concerns about opaque contractual language, in which vague terms such as ``CCTV'' or ``security systems'' could silently encompass AI-driven smart home devices (e.g., cameras with sound detection) without explicitly naming their AI capabilities or data retention policies. P16 argued that such vagueness widened the gap between DWs' expectations and actual practices: \textit{``It's harder to push back against AI monitoring during performance disputes [...] or to question why this AI footage is still kept after the job ends.''} Similarly, P7 proposed concrete contractual improvements, emphasizing the need for clear rules governing data practices during and after employment: \textit{``Perhaps implementing a contract where individuals consent to being recorded could be beneficial; once employment ends, all recorded videos must be erased. There could also be provisions for individuals to indicate whether they want their videos to be retained by the employers after leaving their employment.''} Participants further noted that contracts rarely specified the types of data being collected (e.g., motion-detection data) or how data would be used, stored, or shared.

\begin{tcolorbox}[colback=green!10, colframe=black!20, boxrule=0pt, sharp corners=all]
\textbf{Key Takeaways:} DW agencies emerged as hidden threat actors. Participants reported that agencies prioritized employers’ interests while offering little privacy training or guidance. Employment contracts normalized surveillance practices without clarifying data flows among employers, agencies, and third parties. By failing to provide advice on data flows or support for employer–employee negotiations, agencies effectively acted as institutional co-owners of privacy rules—extending employers’ surveillance reach and constraining DWs’ ability to enforce privacy boundaries.\end{tcolorbox}

\subsection{Ownership and Control in Own Homes} \label{own_home}
\textbf{Exercising Ownership: Configuring Devices and Protecting Others.} \label{exercising_ownership}
Most participants owned and controlled access to their own devices (e.g., by setting up a super-administrator profile and configuring core settings). P3 described their privacy boundary considerations when deploying a smart home device with AI-detection features in their parent’s home for caregiving purposes: \textit{``The camera is positioned in areas where it doesn't intrude on their private spaces or their neighbor's. Mostly, the camera is off and on standby, but it will activate when it detects people.''} In contrast to employer-controlled homes, participants generally characterized their own homes as spaces where they could decide which devices to install, how they were configured, and who would be recorded or listened to.

Moreover, participants leveraged their administrative control to establish clearer privacy boundaries in their own homes—both to protect their own interests and to actively safeguard bystanders—rather than merely enduring surveillance decisions made by others. P7 illustrated this agency: \textit{``The delivery company left a parcel outside during heavy rain. The package was in a cardboard envelope, not even plastic, so it got soaked. So, we used the footage that served as evidence to show the company that it wasn't right.''} Some participants also reported informing visitors about camera placement or device settings, while others adjusted camera orientation to avoid capturing neighbors. Reflecting this heightened privacy awareness and empathy toward those outside the household, P5 noted their efforts to account for neighbors’ privacy when configuring devices.

\textbf{Gendered Admin Roles and ``Passenger'' Users.} In contrast to the significant power imbalances and privacy turbulence participants experienced with employers and DW agencies in workplace contexts, they reported far less turbulence at home, where they generally had greater autonomy and stronger privacy ownership. However, gender disparities emerged clearly in household decision-making around smart home technologies. In many households, men occupied the primary roles as information senders and controllers, managing device configurations and permissions, while women were more often positioned as subjects of data collection, assuming ``passenger user'' roles in multi-user smart homes. P7, for example, highlighted how gendered power dynamics shaped device control, noting that the household admin (their father) retained exclusive authority to manage, delete, or share recorded data: \textit{``My dad bought it and has the only admin permissions. But I think it's more of a cultural thing. He sees himself as the head of the family, so he believes he should have control over the cameras.''} By contrast, P8 emphasized how family values informed privacy ownership boundaries in their home, foregrounding mutual trust and explicit consent over unilateral control: \textit{``Unlike an employment relationship, ours is built on mutual respect and trust. We might glance at each other through the camera sometimes, but secretly monitoring each other goes against what we value. We just stay close matters more than surveillance.''}

\textbf{Ambivalent AI Adoption at Home: Partial Use, Avoidance, and Misunderstandings.} 
Most participants reported limited technical understanding of how AI-enabled devices handled data in their own homes. This was especially evident when discussing features such as voice activation, motion detection, and cloud-based recording, which were often experienced as ``black box'' functions that were difficult to verify, audit, or fully disable. However, prior encounters with opaque workplace surveillance made participants more cautious and configuration-conscious at home. Several participants, like P17, described proactively adjusting the settings of newly purchased cameras by disabling features such as smart motion detection to avoid replicating the feeling of being constantly watched at work. P17 also highlighted the ongoing financial burden of AI add-ons, noting that beyond the one-time device cost, premium analytics and cloud storage required recurring payments.

While some participants acknowledged the effectiveness of smart doorbells and cameras, they expressed limited trust in Alexa-like voice assistants. Even when listening features were turned off for privacy, devices sometimes activated unexpectedly, undermining confidence in control. Several worried that they \textit{``might be doing privacy wrong at home,''} assuming that turning off a setting or closing an app would stop all data collection while still suspecting that devices might listen or store data in the background. Uncertainty about data retention, storage locations, and third-party access further eroded this sense of control. For example, P15 described avoiding always-listening assistants altogether, or placing voice devices only in shared areas, as they \textit{``didn't want to repeat that feeling from work where every word could be saved somewhere. In my own bedroom, I don't want any device that might be recording without me noticing.''}

\begin{tcolorbox}[colback=green!10, colframe=black!20, boxrule=0pt, sharp corners=all]
\textbf{Key Takeaways:} At home, DW participants shifted from passive subjects to active device administrators, managing settings and access to protect family members and bystanders. However, these roles were often dominated by male relatives. Past exposure to opaque workplace surveillance led some DWs to disable AI features, while limited technical knowledge, unexpected device activations, and ongoing subscription costs contributed to uncertainty about actual practices.\end{tcolorbox}

\subsection{How Work and Home Experiences Shaped Each Other} \label{work_home_reshape}

\textbf{Work to Home: Surveillance-Induced Vigilance and Stricter Privacy Configurations at Home.} Many participants described distinct—and often contrasting—privacy challenges across work and home contexts, noting that experiences in employers' homes frequently influenced how they configured and used devices at home. In their personal spaces, participants often acted as primary agents of privacy, managing their own data flows, configuring devices, setting access permissions, and deciding what data (e.g., video feeds) would be stored, shared, or deleted. Several participants reported that experiences with surveillance in employer households—particularly involving AI-driven smart home devices—prompted them to adopt stricter privacy practices in their own homes.

For example, after realizing how employers could monitor and store video or audio without clear consent, P13 described reconfiguring home devices by adjusting doorbell placement, posting stickers on the front door to notify passersby, and limiting facial recognition features. Similarly, some participants disabled cloud backups or turned off certain “smart” options at home after witnessing how easily such data could be shared or replayed in employer households. P15 explained that they disabled camera facial recognition and restricted access permissions to themselves alone, emphasizing that observing employers share footage without notice made them far more cautious at home: \textit{“What I experience at work influences where I install devices and how I use them at home. If I don’t like how they watch me there, I won’t do the same to people in my house.”}

\textbf{Home to Work: Owner Expertise, Expectations, and Frustrations.} This influence also ran in the opposite direction: participants who gained experience configuring and managing devices at home developed clearer expectations for how responsible monitoring should be conducted in employer homes. Many noted that the skills and confidence they acquired in managing their own devices shaped their expectations of how employers ought to handle monitoring. Yet, as P17 described, these expectations were often unmet in practice, leading to frustration and a sense of powerlessness: \textit{“Once we knew that things like doorbells, cameras, and speakers could have notices, access limits, or delete options at home, it felt even harder to accept the vague, take-it-or-leave-it kind of surveillance at work.”}

Most participants viewed privacy as crucial in both contexts, recognizing that their home experiences shaped their comfort levels and expectations when navigating the very different context of employer-controlled spaces. P16 emphasized this interplay: \textit{“My privacy boundaries at home influence what I expect in terms of safety and respect at work [...] privacy in an employer's home goes beyond just personal information. It's also about respecting personal boundaries and ensuring they are not violated.”} Despite these challenges, most participants did not actively seek external advice or guidance in either context. Their strategies were largely intuitive, shaped by personal experiences and prior knowledge from home, which often proved insufficient to navigate the more complex surveillance challenges encountered at work.

\textbf{Mitigation Practices Between Home and Work Contexts.} Between both home and work contexts, most participants relied on a repertoire of mitigation practices that traveled with them, adapting to the distinct power dynamics in each setting. Some described actively managing privacy intrusions by being cautious about their actions—hiding their faces or turning their backs to monitoring cameras, controlling their voice volume, and avoiding personal conversations near cameras and speakers. At home, similar strategies—such as turning off devices when guests visited, repositioning cameras, or limiting which rooms contained smart devices—could be implemented unilaterally. To cope with AI-driven smart cameras, a few participants described behaviors like P7’s: \textit{“I would deliberately position myself with my back to the AI camera because I felt uncomfortable being constantly watched. It was the only moment I could have some privacy, so I would use that time to text or do other activities.”}

Some participants also discussed negotiating camera settings with employers as a workplace mitigation strategy. P3 shared: \textit{“I asked them to move the camera from the bedroom to the living room. I had a place at first where even the bathroom was being covered. So, I complained about it, and something was done about it—telling them your concerns sometimes works.”} Conversely, a few participants faced resistance when requesting privacy. Some employers refused to rotate cameras due to security and safety concerns and did not fully understand participants’ privacy needs (e.g., P11). P10 emphasized the importance of clear communication and mutual respect in addressing workplace privacy: \textit{“It’s the only way we can work towards achieving equality in such situations. But some employers were accommodating, while others prioritized their home security over your privacy.”}

\begin{tcolorbox}[colback=green!10, colframe=black!20, boxrule=0pt, sharp corners=all]
\textbf{Key Takeaways:} Workplace surveillance made DW participants more cautious at home, leading them to adjust device settings and insist on greater transparency. In turn, their experience managing devices at home increased their awareness of—and frustration with—employers’ failure to provide similar accommodations. Across both contexts, participants relied on shared coping strategies, yet they had far less agency in employer homes, where power imbalances constrained their ability to influence surveillance practices.
\end{tcolorbox}
\section{Discussion}

\subsection{Rethinking Threat Models and Pathways}

\subsubsection{A CPM-Grounded Socio-Technical Threat Model} We present a socio-technical threat model grounded in our analysis using a CPM lens~\cite{petronio2017communication,petronio2020conceptualization}. Across workplace and home contexts, our findings reveal repeated instances in which DW participants’ desired privacy boundaries were sidelined by employers, DW agencies, or family members with greater structural or technical power, aligning with prior work calling for threat models that foreground structural risks, lived practices, and uneven control over privacy rules rather than abstract notions of “users” and “attackers”~\cite{warford2022sok,bellini2024sok}. We identify key assets requiring protection, including sensitive data (e.g., audio and video), embodied routines and movements at work, conversational and relational content, reputation and job security, private-life signals that bleed into work (e.g., family calls, health, emotions), and authority over rules governing visibility and data use. Adversaries and threat actors include employers as primary device owners and administrators, DW agencies as intermediaries, technically privileged household members, and platform ecosystems that store and surface AI-generated outputs such as alerts and logs. Threats arise when AI-driven smart home devices obscure privacy ownership, weaken participants’ ability to enforce rules, and trigger privacy turbulence through automated detection and classification, higher-level inference, and persistent data retention across contexts—conditions exacerbated by power imbalances, opaque AI functions, and limited access to settings or data deletion—resulting in harms such as over-monitoring, behavioral misinterpretation, retaliation risks, unauthorized sharing, and persistence of personal data beyond employment.

\subsubsection{Increasing Threats from AI-Driven Smart Home Devices}
In employer-controlled smart homes, AI-driven smart cameras and speakers introduce threat pathways that extend beyond those associated with traditional smart home devices. Prior work has shown that people often underestimate how much data smart home devices collect, how long it is retained, and who can access it~\cite{lau2018alexa,kraemer2020further}. Our participants confirmed these findings, but under conditions of employment where misconfigurations and data over-collection carried significantly higher stakes. In such environments, AI-enabled systems learn from behavior, infer routines, and maintain long-term logs of video, voice, and interaction histories, often without transparent disclosure of what data is collected, how long it is stored, or how it may be reused~\cite{staab2023beyond,yao2024survey}.

In the workplace context, AI-driven smart home devices were not merely tools but part of an infrastructure capable of monitoring, documenting, and potentially justifying punishment. These devices amplify privacy risks, deepen power imbalances, and increase both privacy turbulence and personal harm for DWs in employer-controlled homes. This creates a distinct threat model: \textbf{AI does not merely record events but may algorithmically (mis)interpret them as vigilance, attentiveness, or negligence.} Features such as crying detection, motion analytics, and behavioral summaries transform everyday actions into performance metrics and potential evidence. When these traces are replayed, exported, or interpreted outside their original context, DWs’ expectation that their data will be deleted once employment ends may not be upheld. While prior AI security research has highlighted memorization and inference attacks in large models~\cite{carlini2021extracting,staab2023beyond}, our findings reveal a complementary socio-technical pathway: \textbf{harms from AI analytics arise not only from external attackers but also from powerful insiders who can selectively mobilize logs, alerts, and clips to discipline workers or impose sanctions.}

\subsubsection{DW Agencies as Intermediaries Amplifying Risks} \label{threat_discussion_agency}
Drawing on \S\ref{agency-threat}, our participants’ interactions with DW agencies introduced an additional institutional layer to the threat landscape—one that is largely absent from technical threat models. Prior work has shown that DW agencies and brokers occupy a central role in transnational domestic work, mediating employment contracts, assisting with immigration paperwork, and managing placements, often while prioritizing employers’ interests over DWs’ well-being and rights~\cite{anderson2000doing,he2025exploring}. Our participants described agencies in similar terms: entities that controlled recruitment pipelines, drafted contracts, and mediated complaints, yet routinely shared sensitive information without consent and provided little guidance on privacy or security. Information asymmetries between participants and DW agencies resulted in opaque contract terms that obscured surveillance practices, data retention periods, and data-sharing arrangements, leaving participants unaware of how their data might flow between households, intermediaries, and third parties.

Under the UK GDPR (Article 15), employers are required to inform DWs of any monitoring, explain its purpose, and ensure that data is used only for that stated purpose, avoiding excessive surveillance without consent. In addition, the Employment Practices Code mandates that employers notify employees about CCTV usage and its purpose and conduct impact assessments to balance surveillance needs against privacy intrusions.\footnote{\url{https://ico.org.uk/media2/migrated/1064/the_employment_practices_code.pdf}}) These safeguards are particularly salient in sensitive contexts such as DWs’ private living spaces. However, our findings show that these principles are rarely operationalized in practice. Instead, DW agencies often act as institutional “algorithmic managers” of the employment relationship, standardizing surveillance expectations (e.g., presenting cameras as a normal condition of employment) and framing them as non-negotiable, rather than translating legal protections into concrete and enforceable privacy rules for DWs.

Our CPM-grounded threat model highlights DW agencies as key intermediaries that normalize intrusive monitoring, enforce vague or one-sided privacy policies, undermine DWs’ ability to set personal boundaries, and expand the effective attack surface. The lack of transparency in agency-drafted contracts further heightens these risks. Participants explained that most contracts failed to specify which devices would be used, which AI features (e.g., crying detection) were active, or how AI-generated logs and behavioral data would be stored or shared over time. As a result, participants had little understanding of—or control over—how their personal data might propagate across employers, agencies, and third parties. \textbf{Thus, DW agencies not only placed participants into employer-controlled environments with AI-driven smart home devices, but also shaped the conditions under which surveillance became a default and unquestioned feature of domestic work.}

\subsection{Cross-Contextual Effects of Surveillance on DWs' Privacy Boundaries and Behavior}

Our findings show that DWs' privacy boundaries are continually reshaped by surveillance practices across both employer-controlled and personal smart environments. When DWs are monitored without explicit consent or clear disclosure—such as cameras in bedrooms or AI-enabled tracking they have not agreed to—their ability to establish and enforce boundaries is undermined, prompting ongoing attempts to renegotiate privacy rules that are often unsuccessful. This blurring of boundaries aligns with Vitak et al.'s argument that pervasive surveillance in intimate spaces collapses distinctions between public and private~\cite{vitak2023boundary}. The architecture and placement of cameras and AI-driven smart home devices create a panoptic effect~\cite{foucault2020panopticism, lei2023protecting}, where DWs assume they are always being observed and adjust their behavior accordingly~\cite{johnson2020beyond,he2025exploring}. Participants described acting ``for the camera'' rather than for themselves, reporting anxiety, hypervigilance, and, in some cases, long-term psychological distress.

Our results also highlight that surveillance in employer homes extends beyond those spaces, influencing DWs' privacy practices in their own homes. For example, experiences with opaque monitoring using AI-driven smart speakers that retained prior workers' data prompted many participants to adopt stricter configurations at home. After witnessing employers store or share footage without consent, participants described disabling advanced functions such as motion detection, restricting camera feed access, and using guest modes or warning stickers for visitors. These proactive measures illustrate how participants, once in positions of administrative control, applied lessons from surveillance-heavy workplaces to protect their families and bystanders. Consequently, oppressive monitoring in employer-controlled households produced not only privacy turbulence at work but also heightened vigilance and more deliberate boundary management in participants' personal spaces.

Notably, the influence also ran in the opposite direction. As device owners or co-owners at home, participants gained practical experience configuring privacy rules (e.g., disabling microphones, adjusting camera angles, turning off devices for guests) and developed clearer expectations of what responsible surveillance should look like in employer-controlled homes. Participants who had successfully negotiated respectful monitoring norms within their own households reported feeling particularly unsettled when confronted with ``take it or leave it'' surveillance at work, where they had little control over placement, retention, or sharing.

Finally, our findings show that domestic arrangements themselves are shaped by power and gender: male relatives often acted as sole administrators, leaving participants as ``passenger users'' dependent on others to adjust settings, consistent with prior studies~\cite{kraemer2020further,strengers2019protection,albayaydh2023examining,geeng2019s,despres2024my,he2025exploring}. This underscores that privacy turbulence is not limited to workplaces but occurs whenever individuals with greater social or technical authority define privacy rules for others. We argue, therefore, that DWs' privacy perceptions and practices are not confined to single settings but are continuously negotiated across overlapping homes, institutions, and ecologies: \textbf{experiences in one context recalibrate expectations and behaviors in another, as privacy and safety risks are shaped by a network of homes, agencies, and platforms.}

\subsection{Implications and Suggestions} \label{design_recommendations}

\subsubsection{Interventions for DW Agencies and Policymakers}

Many participants demonstrated limited awareness of their employment rights and entitlements, particularly regarding data protection and privacy. While some contractual terms were verbal, clear written statements of conditions are essential—especially for DWs whose immigration status depends on continued employment~\cite{ahlberg2022the}. As discussed in \S\ref{threat_discussion_agency}, weak regulations and limited oversight allow DW agencies to exploit legal and informational gaps, exacerbating existing power imbalances. Our findings suggest three key directions:

First, DW agencies should adopt transparent and fair employment practices, echoing principles reflected in the UK GDPR. This includes clearly specifying whether and where AI-driven smart home devices are used, what types of information may be collected, how data will be stored and for how long, and under what conditions it may be shared or deleted. Ideally, such information should be provided in DWs' first languages to ensure informed consent~\cite{bernd2022balancing,albayaydh2022exploring,slupska2022they}.

Second, DW agencies should actively advocate for DWs’ privacy interests rather than serving solely as client-facing intermediaries. This involves supporting DWs in negotiating privacy-protective terms with employers, intervening when surveillance is excessive or intrudes into private spaces, and providing ongoing assistance in dispute resolution. Dedicated training and educational initiatives can enhance DWs’ understanding of their legal rights, common surveillance practices (particularly AI-enabled monitoring), and effective negotiation strategies. Delivering these trainings in accessible formats, such as short videos~\cite{he2025exploring} or AI-based advisory modules~\cite{albayaydh2024innovative}, can reduce barriers to engagement and accommodate DWs’ limited availability.

Third, DW-specific privacy protections should be integrated into the Employment Practices Code, drawing on international frameworks such as the ILO DW Convention (No. 189).\footnote{Domestic Workers Convention (No. 189) establishes fundamental rights and standards for DWs, including provisions for decent living conditions and privacy for live-in DWs (Articles 6 and 17): \url{https://normlex.ilo.org/dyn/nrmlx_en/f?p=NORMLEXPUB:12100:0::NO::P12100_ILO_CODE:C189}. The UK has not ratified this convention.} Such standards should prohibit surveillance in highly intimate spaces (e.g., bathrooms, bedrooms) and require proportionality assessments that account for DWs’ dependence and embeddedness in employers’ households.

\subsubsection{Design Implications for AI-Driven Smart Home Devices}

Drawing on our findings in \S\ref{pervasive_surveillance}, \S\ref{AI_threat}, and \S\ref{own_home}, AI-driven smart home devices should provide clear, continuous indicators when recording or analyzing data, rather than relying on ambiguous LEDs or app-only notifications~\cite{slupska2022they,zeng2017end,albayaydh2023examining,thakkar2022would,albayaydh2024innovative}. These indicators should be understandable to all household members, including DWs and visitors, and clearly reflect when privacy or guest modes are active. Such interventions should not rely solely on voluntary adoption. Robust guest or bystander mode designs should be both auditable and accessible: DWs should be able to see whether AI analytics are active, what data is retained, and when settings are changed, even if they cannot directly modify configurations. For example, indicators could include: (i) an audible cue when analytics are enabled or continuous processing begins, and (ii) an in-home display showing active sensors and AI functions across devices. Crucially, these indicators should be glanceable without app access and resistant to tampering by uncooperative or malicious employers. Furthermore, smart home devices should include configurable guest and bystander modes that temporarily restrict data capture, disable AI analytics, or limit retention during certain activities or periods. Building on DWs’ own practices at home, these modes should be easy to activate, supporting co-ownership of privacy rules across multiple stakeholders~\cite{geeng2019s,lau2018alexa,albayaydh2023examining}.

To address concerns about residual traces and orphaned data, smart home devices should implement automatic deletion or strong de-identification of behavioral data when employment relationships end, or when a user or bystander profile is removed~\cite{kim2022identification}. At minimum, data associated with specific DW profiles should be anonymized to prevent retrospective monitoring or discipline of former workers. Additionally, smart home ecosystems should support controls that allow non-admin household members, including DWs, to inspect what data about them has been collected, request deletions, and set personal retention preferences~\cite{saqib2025bystander,h2022monitoring,lau2018alexa,albayaydh2023examining}. These controls can shift smart homes toward genuine multi-user environments rather than reinforcing single-admin models that mirror existing power hierarchies.

Finally, as LLMs are increasingly embedded in AI-driven smart home devices mediating children’s everyday interactions, new risks arise: these systems can hallucinate, misinterpret, or infer sensitive information, leading to unintended data exposure and lack of accountability~\cite{jiao2025llms,staab2023beyond,mireshghallah2023can}. To mitigate these risks, devices should avoid turning domestic childcare workers into unprotected content moderators or data stewards by inspecting children’s interactions with devices (e.g., prompts children provide to smart speakers). Instead of exposing raw transcripts that could be misused, platforms could employ privacy-preserving, interpretability-aware interfaces that flag only problematic patterns while maintaining explicit boundaries on data access and retention. By aligning technical affordances with DWs’ privacy boundaries in both work and home contexts, designers and policymakers can better support privacy as an everyday, negotiated practice~\cite{he2025privacy}.
\section{Conclusion}
We interviewed 18 DW participants in the UK to examine how they experienced and managed privacy with AI-driven smart home devices in both employer-controlled households and their own homes. Using CPM as our analytic lens, we showed that participants’ privacy boundaries were continually renegotiated in response to camera surveillance, AI analytics, and agency-mediated employment arrangements. In employer homes, smart cameras and speakers extended monitoring from simple recording to algorithmic judgment and long-lived behavioral traces, producing panoptic effects and residual risks that persisted beyond employment. We also identified DW agencies as institutional threat actors, whose opaque contracts, limited training, and employer-sided dispute handling normalized intrusive surveillance and constrained workers’ capacity to contest monitoring. In their own homes, participants were not uniformly vulnerable bystanders but device owners and privacy guardians for children, parents, and guests. Experiences of coercive surveillance at work informed stricter configurations and more cautious AI adoption at home, while home-based administrative experiences sharpened expectations and sometimes led to frustrations in employer settings. We synthesize these cross-context dynamics into a CPM-grounded socio-technical threat model highlighting DWs’ boundary work and specifying assets, adversaries, and harms in AI-driven smart home ecologies. Overall, our findings underscore that protecting DWs’ privacy requires coordinated changes to device and platform design, regulatory safeguards, and agency practices—not solely individual coping strategies.


\section*{Acknowledgments}
We thank our participants for their involvement in this study and the anonymous reviewers for their valuable feedback. We also acknowledge the use of ChatGPT for assistance with proofreading and grammar checking.



\bibliographystyle{IEEEtran}
\bibliography{reference}
%



\appendix

\section{Participant Demographics} \label{demographic}
\renewcommand{\arraystretch}{1}
\begin{table*}[!ht]
    \centering
    \fontsize{10pt}{10pt}\selectfont
    \resizebox{1\textwidth}{!}{%
    \begin{tabular}{lllllp{1.5cm}lccp{3cm}p{3cm}}
    \toprule
        \textbf{ID} & \textbf{Age} & \textbf{Gender} & \textbf{Education} & \textbf{Ethnicity} & \textbf{Job} & \textbf{Experience} & \textbf{Agency} & \textbf{Migrant?} & \textbf{Workplace devices} & 
        \textbf{Owned devices} \\ 
    \midrule
        P1 & 35-44 & Woman & Some College & Asian & Nanny & 6-10 yrs. | Live-in & Yes & Yes & Smart home camera, Baby cam/monitor, Smart doorbell, Smart TV, Smart speaker, Smart doorlock & 
        Smart TV, Smart speaker\\
    \hline
        P2 & 25–34 & Man & Assoc./Tech.  & Black & House cleaner, housekeeper & 1-3  yrs. | Live-out & Yes & Yes & Smart home camera, Smart doorbell, Smart TV, Smart speaker, Smart doorlock & 
        Smart home camera, Smart doorbell\\
    \hline
        P3 & 25–34 & Man & Assoc./Tech.  & Black & Caregiver & 6-10 yrs. | Live-in & Yes & No & Smart home camera, Smart TV, Smart speaker & 
        Smart TV, Smart speaker, Smart doorbell\\
    \hline  
        P4 & 25–34 & Woman & Master & White & Nanny & 6-10 yrs. | Live-out & Yes & Yes & Smart home camera, Smart speaker & 
        Smart speaker\\ 
    \hline 
        P5 & 18–24 & Woman & Bachelor & Asian & Caregiver & 1-2 yrs. | Live-in & Yes & No & Smart home camera, Smart speaker, Smart doorlock & 
        Smart doorbell\\
    \hline 
        P6 & 25–34 & Woman & Master & White & Babysitter, pet carer & 6-10 yrs. | Live-out & No & No & Smart home camera, Smart doorbell, Smart TV, Smart speaker, Smart doorlock & 
        Smart home camera, Smart doorbell, Smart speaker\\
    \hline 
        P7 & 18–24 & Woman & Assoc./Tech.  & Asian & Babysitter, nanny & 1-2 yrs. | Live-in & No & Yes & Baby cam/monitor, Smart TV, Smart speaker, Smart doorlock & 
        Smart TV, Smart home camera\\
    \hline 
        P8 & 25–34 & Woman & Some College & Black & Babysitter, house cleaner & 3-5 yrs. | Live-out & No & Yes & Smart home camera, Baby cam/monitor, Smart doorbell, Smart TV, Smart speaker, Smart doorlock & 
        Smart home camera\\
    \hline 
        P9 & 25–34 & Woman & Bachelor & Asian & House cleaner, caregiver & 1-2 yrs. | Live-out & No & No & Smart home camera, Smart doorbell, Smart doorlock & 
        Smart TV, Smart speaker\\
    \hline 
        P10 & 35–44 & Woman & Assoc./Tech.  & White & Caregiver & 6-10 yrs. | Live-in & Yes & Yes & Smart home camera, Smart TV, Smart speaker, Smart doorlock & 
        Smart TV, Smart speaker, Smart doorbell\\
    \hline 
        P11 & 25–34 & Man & High school  & Black & Gardener, house cleaner & 1-2 yrs. | Live-out & No & No & Smart home camera, Smart TV, Smart speaker & 
        Smart TV, Smart home camera, Smart doorbell\\
    \hline 
        P12 & 25–34 & Woman & Assoc./Tech.  & Black & Caregiver & 1-2 yrs. | Live-out & No & Yes & Smart home camera, Smart TV, Smart speaker & 
        Smart doorbell\\
    \hline 
        P13 & 55+ & Man & High school  & White & Maintenance worker & 10+ yrs. | Live-out & No & No & Smart home camera, Smart doorbell & 
        Smart TV, Smart doorbell\\
    \hline 
        P14 & 35-44 & Man & Some College & Asian & House cleaner & 1-2 yrs. | Live-out & No & Yes & Smart home camera, Smart speaker, Smart doorbell & 
        Smart home camera, Smart TV, Smart speaker\\
    \hline 
        P15 & 25–34 & Woman & Bachelor & Black & Nanny & 3-5 yrs. | Live-in & No & Yes & Smart home camera, Baby cam/monitor, Smart doorbell, Smart TV, Smart speaker, Smart doorlock & 
        Smart home camera, Smart doorbell, Smart TV, Smart speaker\\
    \hline 
        P16 & 25–34 & Man & Assoc./Tech.  & Asian & Caregiver & 5-10 yrs. | Live-in & Yes & Yes & Smart home camera, Smart doorbell, Smart TV, Smart speaker, Smart doorlock & 
        Smart home camera\\
    \hline 
        P17 & 35-44 & Man & Some College & Asian & House cleaner, housekeeper & 5-10 yrs. | Live-out & No & No & Smart home camera & 
        Smart home camera, Smart doorbell\\
    \hline 
        P18 & 18–24 & Woman & Bachelor & White & Pet carer & $<$1 yr. | Live-out & No & No & Smart home camera, Smart doorbell & 
        Smart doorbell, Smart speaker\\
    \bottomrule 
    \end{tabular}}
    \caption{Participant Demographics and Details ($N=18$).}
    \label{tab:demographic}
\end{table*}





\end{document}